\documentclass[aps,prl,preprint,groupedaddress,amsmath,amssymb,showpacs]{revtex4}

\usepackage{graphicx}
\usepackage{dcolumn}
\usepackage{bm}

\def\const{\mathop {\rm const}\nolimits } 
\def\tr{\mathop {\rm tr}\nolimits } 
\begin{document}


\title{Tension-induced multistability in inextensible helical ribbons}


\author{E.~L.~Starostin}
\email{e.starostin@ucl.ac.uk}
\author{G.~H.~M.~van~der~Heijden}
\email{g.heijden@ucl.ac.uk}
\affiliation{Centre for Nonlinear Dynamics, University College London,\\
 Gower Street, London WC1E 6BT, UK}

\date{\today}

\begin{abstract}
We study the non-monotonic force-extension behaviour of helical ribbons
using a new model for inextensible elastic strips. Unlike previous rod
models our model predicts hysteresis behaviour for low-pitch ribbons of
arbitrary material properties. Associated with it is a first-order transition
between two different helical states as observed in experiments with cholesterol
ribbons. Numerical solutions show non-uniform uncoiling with hysteresis also
occurring under controlled tension. They furthermore reveal a new uncoiling
scenario in which a ribbon of very low pitch shears under tension and
successively releases a sequence of almost planar loops. Our results may be
relevant for nanoscale devices such as force probes.

\end{abstract}


\maketitle



Helical ribbons are common in nature \cite{Galloway02,Smith01} and hold great promise as parts of nano-scale force probe devices \cite{Gao05}. 
Smith et al.~\cite{Smith01} studied the force-response behaviour of helical ribbons that appear as metastable intermediates in the process of
cholesterol crystallisation in the native gallbladder bile. Cantilever tests showed linear, Hookean, and reversible behaviour at low tension,
while at sufficiently high tension the ribbon becomes metastable and separates into a straight and a helical domain. To explain this behaviour
they proposed a phenomenological model for an inextensible strip. Intrinsic curvature entered the elastic energy through an isotropic term that
can be interpreted as describing the difference in surface tension between the ribbon's inner
and outer surfaces. The model is only valid for helical configurations and the spontaneous helical pitch of the ribbon is determined by
anisotropy of the strain tensor, rather than by intrinsic geometry. The model shows multistability with large and small pitch solutions, and
associated hysteresis behaviour. However, recent X-ray diffraction studies in \cite{Khaykovich07} show that these cholesterol ribbons are single
crystals and suggest that the crystalline structure rather than the surfactant layers determine the helical form.

Kessler \& Rabin \cite{Kessler03} considered an elastic rod model and by direct energy minimisation were able to produce non-reversible
(hysteresis) force-extension behaviour for helical springs of small pitch and large ratio of torsional to bending stiffness.
%
%
This behaviour was confirmed analytically by Zhou et al.~\cite{Zhou05} who used a rod model that also allowed for transverse
anisotropy. A single hysteresis cycle was found for large stiffness ratios.
%
%
In \cite{Wada07} dynamical simulations were performed using an extensible isotropic rod model revealing the possibility of several
abrupt changes in the force response of a stretched helix, again in the case of unusually large stiffness ratios (corresponding to negative
Poisson ratios).


We propose a new model of an elastic inextensible strip. Such a strip can only bend; thus, if its unstressed surface is developable, i.e., has
vanishing Gaussian curvature, then it will remain so after deformation. We believe that this model is better suited for crystalline structures. 
We will show that it predicts hysteresis behaviour for realistic values of Poisson's ratio and without requiring material anisotropy.


\begin{figure}
\includegraphics[height=2.4in]{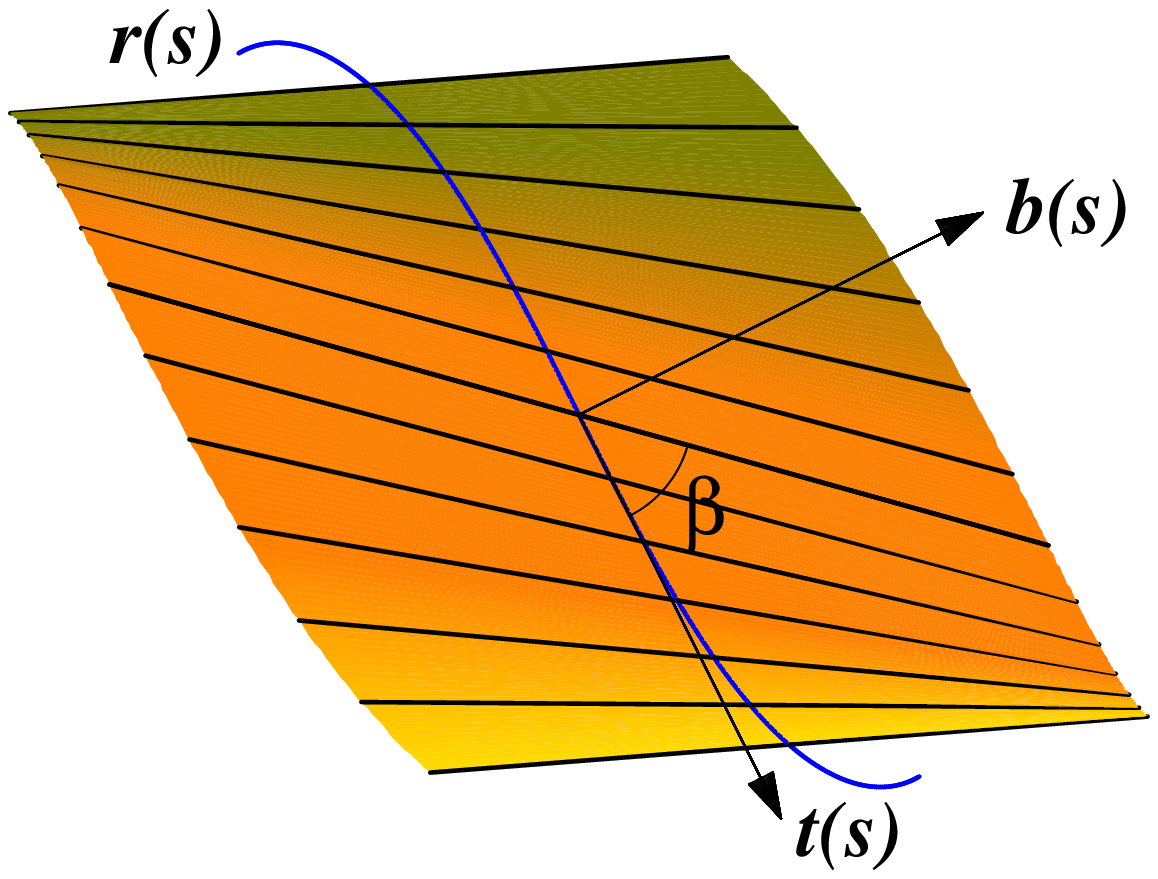}
\includegraphics[height=2.4in]{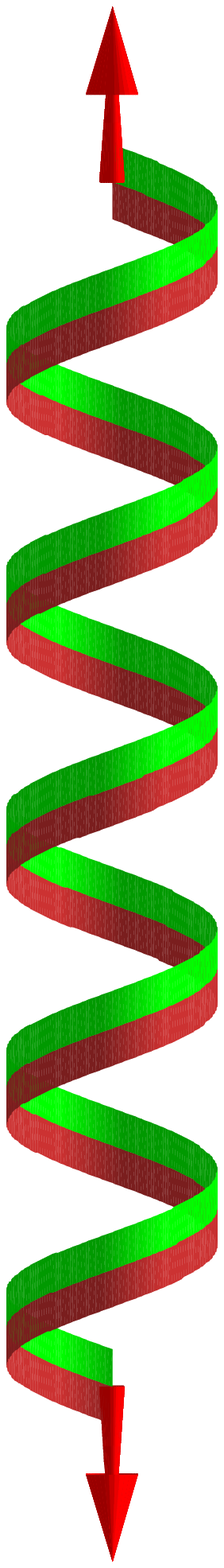}\\
\caption{\label{fig:dvlp_srf} {\it Left:} A developable surface is made up of straight generators at angle $\beta$ to the tangent $\bm t$ to the
centreline $\bm r$; $\bm b$ is the binormal. {\it Right:} A developable helical ribbon under tensile end loading.}
\end{figure}

Consider an inextensible ribbon that, when developed onto a plane, makes a strip of width $2w$ bounded by two parallel straight lines. 
By the property of developability, the strip, however deformed, can be reconstructed from its centreline $\bm r(s)$, which itself is determined, up to Euclidean motion, by its curvature $\kappa(s)$ and torsion $\tau(s)$. Here $s$ is the arclength along the centreline. More precisely, we can parametrise the strip as
\cite{Randrup96}
\begin{eqnarray}
&& \bm x(s,t) = \bm r(s) + t\left[\bm b(s)+\eta(s)\,\bm t(s)\right], \\
&& \tau(s)=\eta(s)\kappa(s), \quad s=[0,L],~t=[-w,w], \nonumber
\label{par_strip}
\end{eqnarray}
where $\bm t$ is the unit tangent vector and $\bm b$ the unit binormal. The generator of the developable surface of the strip makes an angle $\beta=\arctan(1/\eta)$ with the positive tangent direction (Fig.~\ref{fig:dvlp_srf}).

Let $\bm N$ be the unit normal to a smooth surface $\Omega$ and $S(\bm T) = - \partial_{\bm T}\bm N$ the shape operator acting on a unit tangent $\bm T$
to the surface. The bending energy for a Kirchhoff-Love shell of thickness $2h$ can be written as the following integral
over the surface \cite {Neff04,Zorin05,Friesecke03}
\begin{equation}
U=\frac{D}{2}\iint_\Omega  [\nu (\tr(\Delta S))^2 + (1-\nu) \tr((\Delta S)^2)] d\sigma ,
\label{energy1}
\end{equation}
where $D = 2 E h^3/[3 (1-\nu^2)]$ is the flexural rigidity, $\nu$ is Poisson's ratio, $E$ is Young's modulus and $\Delta S = S - S_0$; here and later the subscript $0$ refers to the undeformed state. Equation~(\ref{energy1}) can be rewritten in the form \cite{Zorin05}
\begin{equation}
U=D\iint_\Omega [(1+\nu)(\Delta H)^2 + (1-\nu) ((\Delta A)^2+4 A A_0 \sin^2\theta)] d\sigma ,
\label{energy2}
\end{equation}
where $\Delta H = H - H_0$, $H$ is the mean curvature, $\theta$ is the angle between the principal curvature axes in the deformed and undeformed states, and
$A=\sqrt{H^2-G}$, $G$ being the Gaussian curvature.

For developable surfaces, $H=A=\kappa_1/2$ and $K=0$ ($\kappa_1$ and $\kappa_2\equiv0$ are the principal curvatures).
We shall assume the undeformed strip to be helical, with constant pitch angle $\pi/2-\beta_0$, and to lie on the surface of a cylinder of
radius $R_0$ (Fig.~\ref{fig:dvlp_srf}{\it Right}). Then Eq.~(\ref{energy2}) becomes
\begin{eqnarray}
U=\frac{D}{2}\int_0^L \int_{-w}^{w} \left[\kappa_1^2-2\kappa_1\kappa_{1,0}(1+(\nu-1)\sin^2\theta)+\kappa_{1,0}^2\right] d t d s,
\label{energy3}
\end{eqnarray}
with $\kappa_{1,0} = 1/R_0$ and $\theta = \beta - \beta_0$.
On using Eq.~(\ref{par_strip}) the $t$-integration can be carried out analytically, as in the case without intrinsic curvature \cite{Wunderlich62}, and we arrive at
\begin{equation}
U=Dw\int_0^L g(\kappa,\eta,\eta') d s +\const
\label{energy4}
\end{equation}
with
\begin{eqnarray}
&& g(\kappa,\eta,\eta')=\kappa^2\left(1+\eta^2\right)^2 V(w\eta')-\frac{2}{R_0}\kappa\left[1+\eta^2+q(\eta-\eta_0)^2\right],   \\
&& V(w\eta')=\frac{1}{2w\eta'}\log\left(\frac{1+w\eta'}{1-w\eta'}\right), \nonumber
\label{energy5}
\end{eqnarray}
where $q=(\nu-1)/(1+\eta_0^2)$, $\eta_0=\cot\beta_0$ and the prime denotes the derivative with respect to $s$.
Note that for strips with intrinsic curvature ($R_0 < \infty$), equilibrium shapes will depend on the material properties through Poisson's ratio.
Also note that in the limit of narrow strips, $w\eta' \to 0$, we have $V(w\eta') \to 1$ and no derivative enters the integrand in Eq.~(\ref{energy4}).


A different model of a thin and narrow elastic strip may be obtained as a limiting case of the thin anisotropic elastic rod by pushing one of the
principal bending stiffnesses to infinity \cite{Mahadevan93}. In this model the material frame of the rod coincides with the Frenet frame
of its centreline. We shall call it a Frenet rod. Its surface is swept out by the binormal of the centreline and is not obliged to deform isometrically.
If the centreline has intrinsic curvature $\kappa_0$ and intrinsic torsion $\tau_0$ then the elastic energy reads \cite{Kessler03}
\begin{equation}
U_F= \int_0^L \frac{B}{2} (\kappa - \kappa_0)^2 + \frac{C}{2} (\tau - \tau_0)^2  d s,
\label{frod_energy}
\end{equation}
where $B$ and $C$ are the bending and torsional stiffnesses, respectively.
Assuming a rectangular cross-section, we have $B=4Ewh^3/3$, $B/C = (1+\nu)/2$ and Eq.~(\ref{frod_energy}) may be rewritten as
\begin{equation}
U_F= \frac{B}{2} \int_0^L g_F(\kappa,\eta) d s +\const
\label{energy_rod}
\end{equation}
with
\begin{equation}
g_F(\kappa,\eta)=\kappa^2 \left (1 + \frac{2\eta^2}{1+\nu}\right) - \frac{2\kappa}{R_0(1+\eta_0^2)}\left(1+\frac{2\eta_0\eta}{1+\nu}\right).
\label{frod_energy2}
\end{equation}


For both the strip and the rod model energy minimisation is a 1D variational problem cast in a form that is invariant under Euclidean motions. 
Following \cite{Starostin07}, the Euler-Lagrange equations can be immediately written down in the form of six balance equations
for the components of the internal force $\bm{F}$ and moment $\bm{M}$ in
the directions of the Frenet frame of tangent, principal normal and binormal,
$\bm{F}=(F_t,F_n,F_b)^T$, $\bm{M}=(M_t,M_n,M_b)^T$, and two scalar equations:
\begin{eqnarray}
&& \hspace{-1cm} \bm{F}'+\bm{\omega}\times\bm{F}=\bm{0}, \quad
\bm{M}'+\bm{\omega}\times\bm{M}+\bm{t}\times\bm{F}=\bm{0},
\label{eqs_vector} \\
&& \hspace{-1cm} \partial_\kappa g+\eta M_t+M_b=0, \quad
\left(\partial_{\eta'} g\right)'-\partial_\eta g-\kappa M_t=0,
\label{eqs_scalar}
\end{eqnarray}
where $\bm{\omega}=\kappa(\eta,0,1)^T$ is the curvature vector. The equations have $|\bm{F}|^2$ and
$\bm{F}\cdot\bm{M}$ as first integrals. We note that in the variables $(\kappa,\eta,\eta')$ the first equation in
Eq.~(\ref{eqs_scalar}) is algebraic and the second equation becomes so in the limit of a narrow strip ($w=0$) and for the rod
(Eq.~(\ref{frod_energy2})).


We first consider helical solutions, i.e., we set $\kappa=\const \neq 0$ and $\tau=\const$. In addition, we assume that
$\bm F$ and $\bm M$ are constant in the local (Frenet) frame. Then Eq.~(\ref{eqs_vector}) implies $F_n =0$, $M_n=0$ and
\begin{equation}
\kappa(M_t - \eta M_b) = F_b, \quad F_t = \eta F_b .
\label{eqs_hel}
\end{equation}
%
Thus, the vector $\bm F$ is directed along the helical axis ($z$) and $F_t = F \cos\beta$, $F_b = F \sin\beta$.
The magnitude of the force can be obtained from Eqs.~(\ref{eqs_scalar}), (\ref{eqs_hel}):
$F=\sqrt{1+\eta^2}[\kappa\eta\partial_\kappa g - (1+\eta^2)\partial_\eta g]$ (negative for tension).

We assume that no moment acts about the helical axis, 
i.e., the moment is orthogonal to the force: $\bm{F}\cdot\bm{M}=0$. 
Then the first Eq.~(\ref{eqs_scalar}) simplifies to $\partial_\kappa g = 0$.
This allows us to eliminate the curvature and to obtain an explicit expression for the force
as a function of only one variable, for which it is convenient to choose the angle $\beta$.
We furthermore introduce the normalised extension $\zeta = z(L)/L =\cos\beta$ so that $\eta = \zeta/\sqrt{1-\zeta^2}$.
For our functional $g$ we find for the normalised force $f=F R_0^2$
\begin{equation}
f=\frac{(1-\nu)}{2 \sin\beta} [2(1+\nu)\sin 2(\beta - \beta_0) + (1-\nu) \sin 4 (\beta - \beta_0)],
\label{force1}
\end{equation}
while the analogous expression for the Frenet rod functional $g_F$ is
\begin{equation}
f_F = 4\frac{[(1+\nu)\sin\beta_0 \sin\beta + 2 \cos\beta_0 \cos\beta] \sin(\beta - \beta_0)}{\sin\beta [(1+\nu)\sin^2\beta + 2\cos^2\beta]^2}. 
\label{force2}
\end{equation}
This expression coincides with the corresponding formula for the rod with circular cross-section in \cite{Wada07}
after replacing the torsional-to-bending stiffness ratio in the latter with that for a rectangular cross-section.

\begin{figure}
\includegraphics[width=5in]{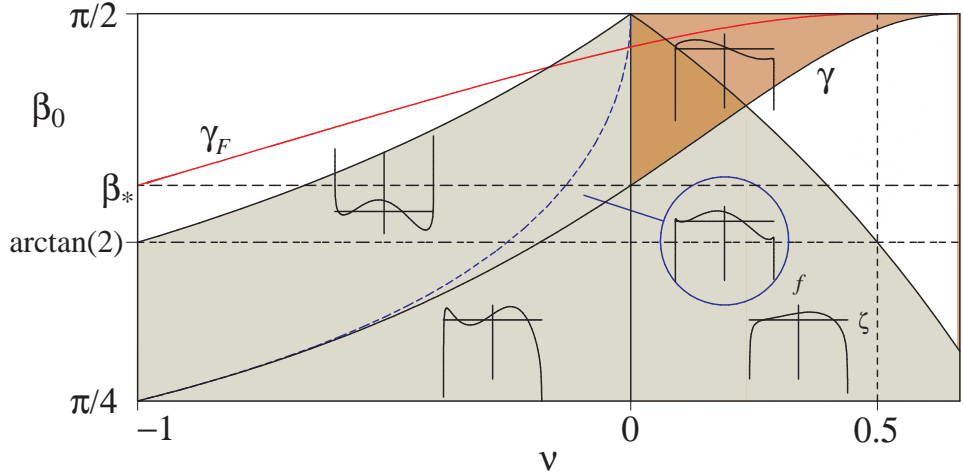}
\caption{\label{fig:spring_const} The $\nu$-$\beta_0$ parameter plane for helical solutions with regions of different force-extension curves Eq.~(\ref{force1}) indicated. The dashed curve is given by $\nu=-\cot^2\beta_0$. Dark shaded is the physical region of multistability. Light shaded is the region, bounded by the curve given by $\cos\beta_0 =|\nu|/\sqrt{1+(1-\nu)^2}$, where the helical solutions are stiffer for the strip than for the rod. $\beta_*=\arctan\left(3/2\right)^{5/2}$.}
\end{figure}

For small deformations $\zeta$, the response of the helical ribbon may be approximated as Hookean (linear) with spring constant 
%
$\left|\left. \partial_\zeta f \right|_{\zeta=\zeta_0}\right| = 4(1-\nu) \sin^{-2}\beta_0$.
%
The corresponding value for the Frenet rod,
%
$\left|\left. \partial_\zeta f_F \right|_{\zeta=\zeta_0}\right|  = 4 [1+\nu + (1-\nu)\cos^2\beta_0]^{-1}$,
%
may be obtained from the classical expression \cite{Love27,Wada07} by replacing the torsional-to-bending stiffness ratio with that for a rod of rectangular cross-section. Which of the two is stiffer depends on parameters, as indicated by the light shading in Fig.~\ref{fig:spring_const}.




It is straightforward to show that for positive $\nu$ and for arbitrary $\beta_0$ the force $f$ vanishes only at two extensions: $\zeta=\zeta_0$ (undeformed state) and $\zeta = \zeta_1<0$, $\zeta_0^2+\zeta_1^2=1$ (helical axis orthogonal to the undeformed flat direction), while for negative $\nu$, $f$ has 4 roots.
Fig.~\ref{fig:spring_const} shows force-extension curves Eq.~(\ref{force1}) as insets in the $\nu$-$\beta_0$ parameter plane. Non-monotonicity of these curves (hysteresis) implies the coexistence of two stable solutions at given force $f$, one of high and one of low pitch. A first-order phase transition between
specific values of these pitches occurs at a certain force for which the total energy densities $g+F\zeta$ of the two helices (`phases') are equal. Solutions may then be constructed composed of arbitrary pieces of each helix. Such solutions describe the phase-separated configurations observed experimentally in \cite{Smith01}. The domain of hysteresis or first-order transitions in the $\nu$-$\beta_0$ parameter plane is bounded by critical curves $\{\nu=0\}$ and $\gamma$ of second-order phase transitions or cusps given by $\partial_{\zeta\zeta} f=0$.
Note that the maximum physically realisable value of Poisson's ratio, $\nu=1/2$, ensures an interval of initial pitches with phase transitions.
The curve $\gamma_F$ in Fig.~\ref{fig:spring_const} is the critical curve for the Frenet rod \cite{Kessler03}, Eq.~(\ref{force2}). It bounds a much smaller region of multistability \endnote{The $11.1^\circ$-pitch ribbon tested in \cite{Smith01} is hysteretic according to the strip model but not according to the rod model.}.


\begin{figure}
\includegraphics[width=5in]{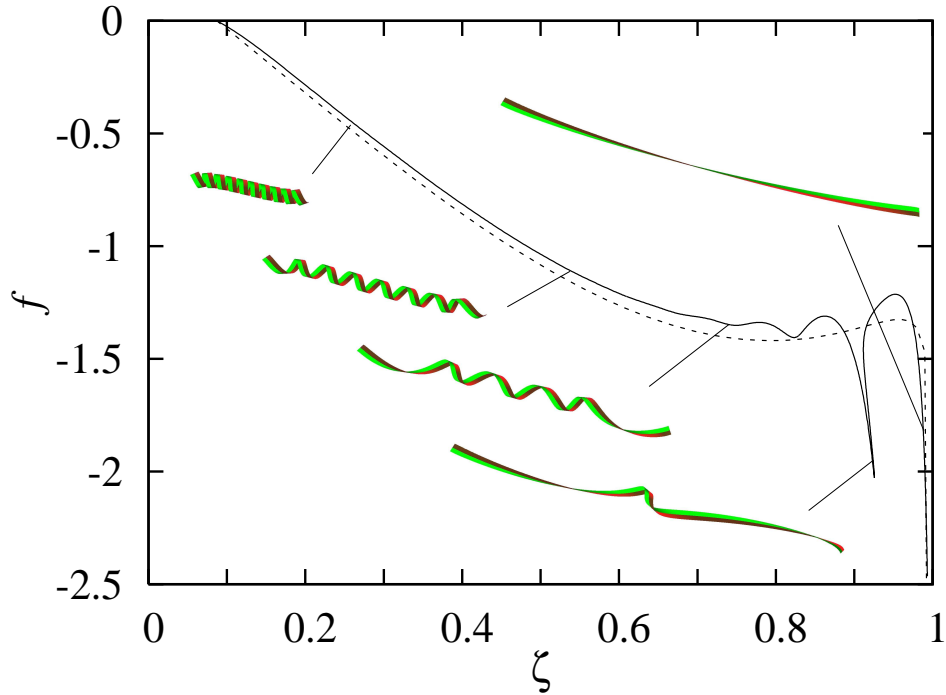}
\caption{\label{fig:h5shapes} Force-extension diagram for the inextensible strip ($n=10$, $\nu=0.3$, $\beta_0=85^\circ$).
The dashed curve corresponds to Eq.~(\ref{force1}) for helical solutions.}
\end{figure}

\begin{figure}
\includegraphics[width=3.15in]{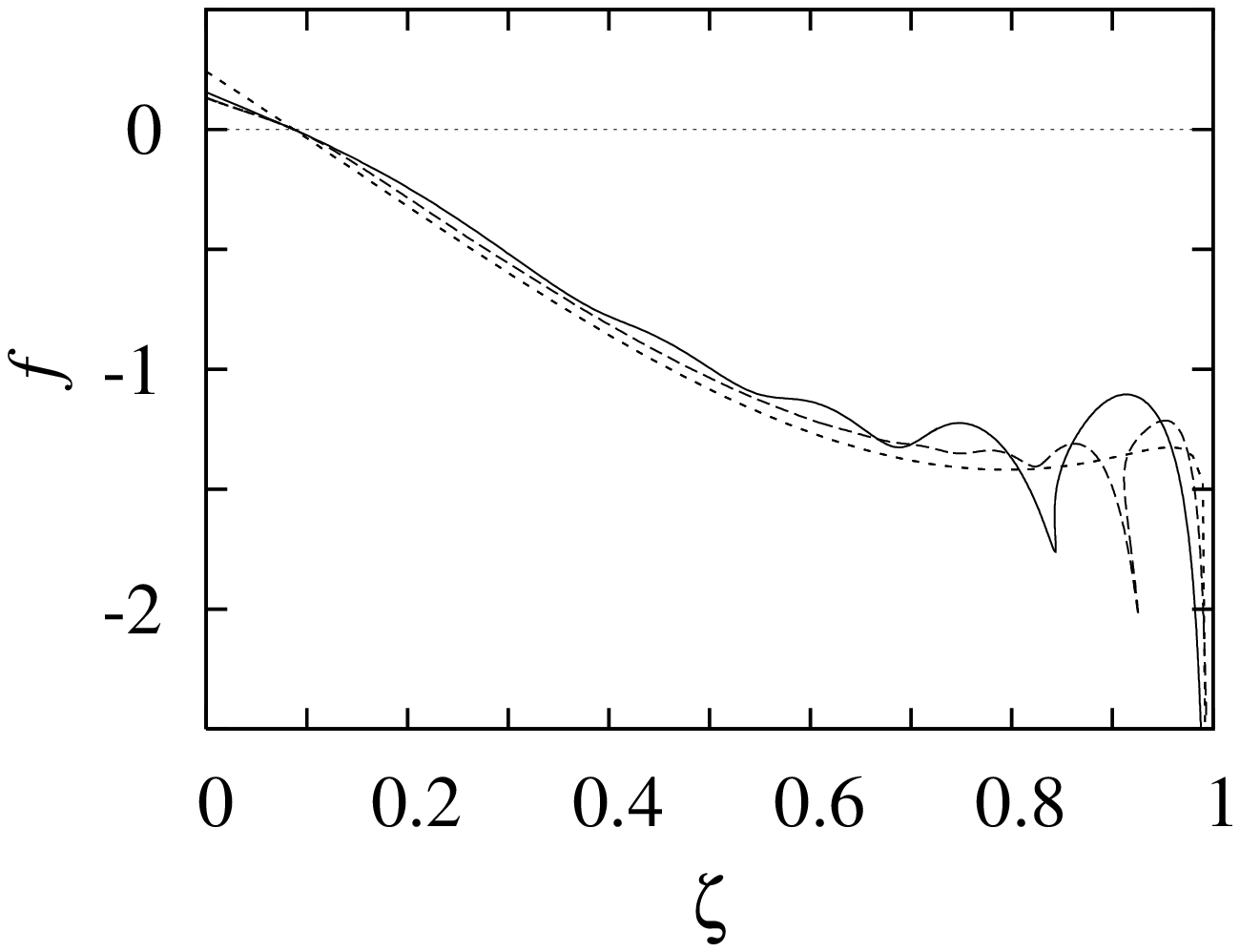}
\includegraphics[width=3.2in]{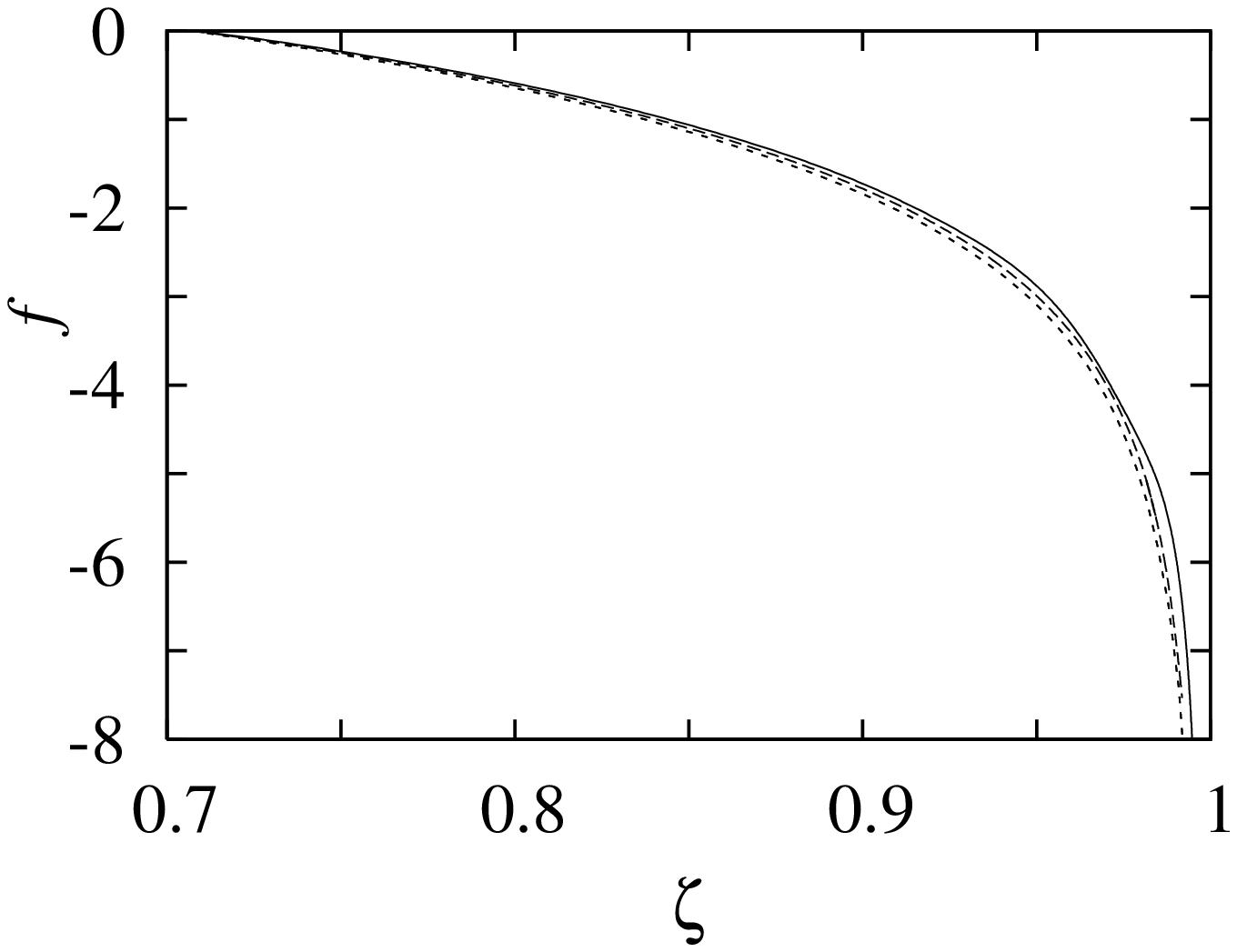}
\caption{\label{fig:h37} Force-extension diagram for the inextensible strip: $n=5$ (solid), $n=10$ (dashed); $\nu=0.3$. {\it Left}: $\beta_0=85^\circ$.   {\it Right:} $\beta_0=45^\circ$. Dotted curves are for exact helical solutions.}
\end{figure}

\begin{figure}
\includegraphics[width=3.2in]{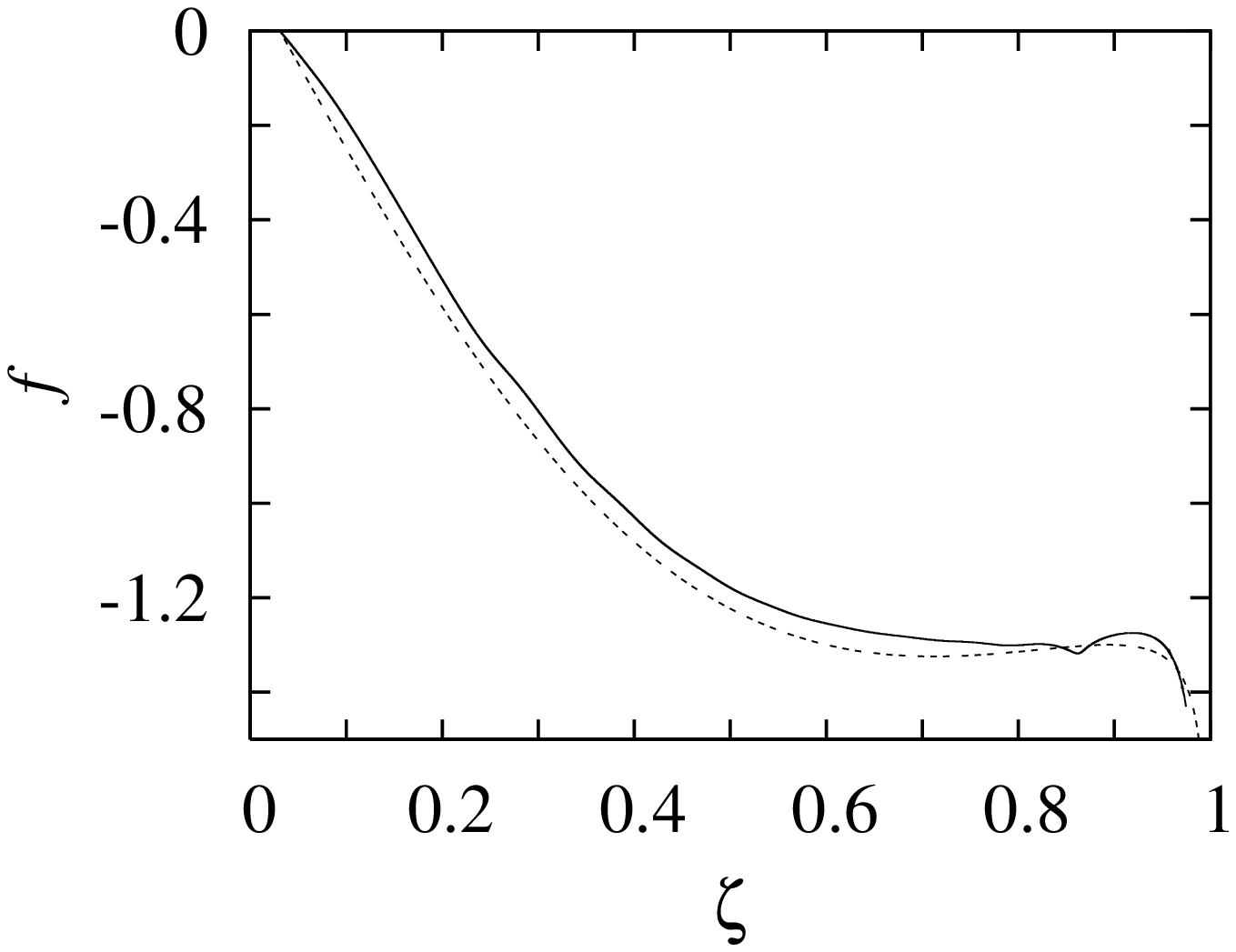}
\includegraphics[width=3.1in]{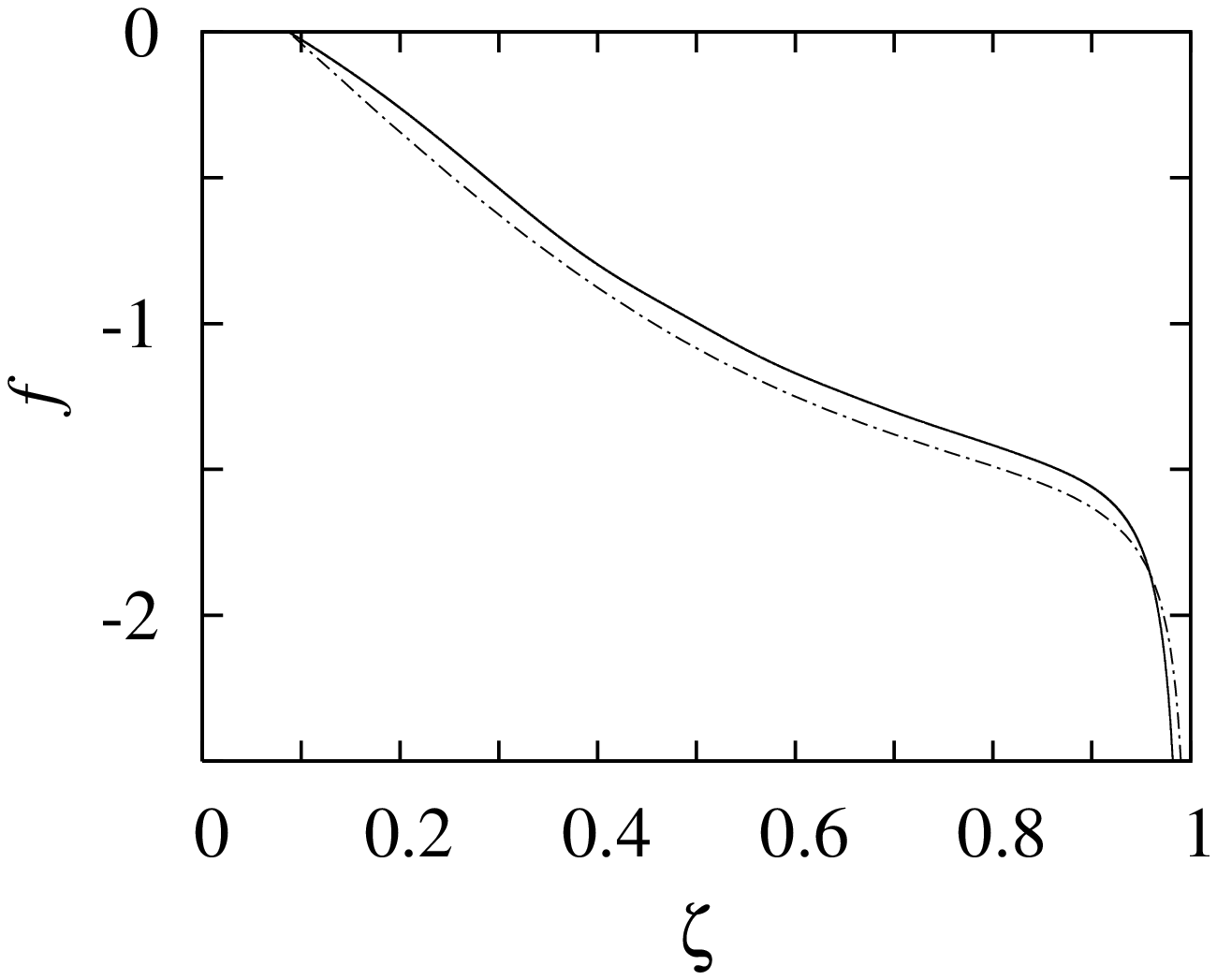}
\caption{\label{fig:h108} Force-extension diagram for the Frenet rod. {\it Left}: $n=10$, $\nu=0.1$,  $\beta_0=88.24^\circ$.   {\it Right:} $n=5$, $\nu=0.3$, $\beta_0=85^\circ$. Dashed curves are for exact helical solutions.}
\end{figure}

\begin{figure}
\includegraphics[width=3.2in]{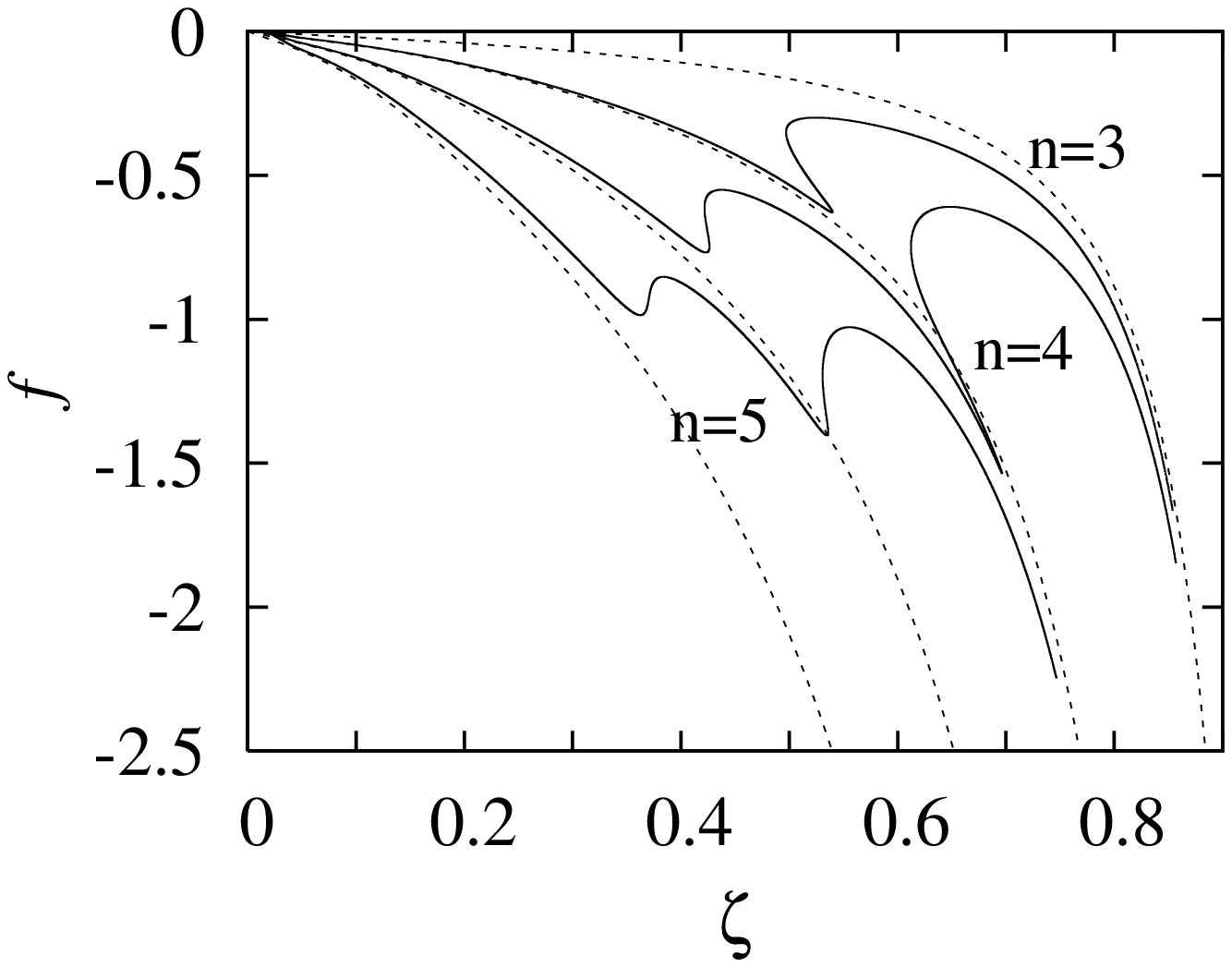}
\includegraphics[width=3.2in]{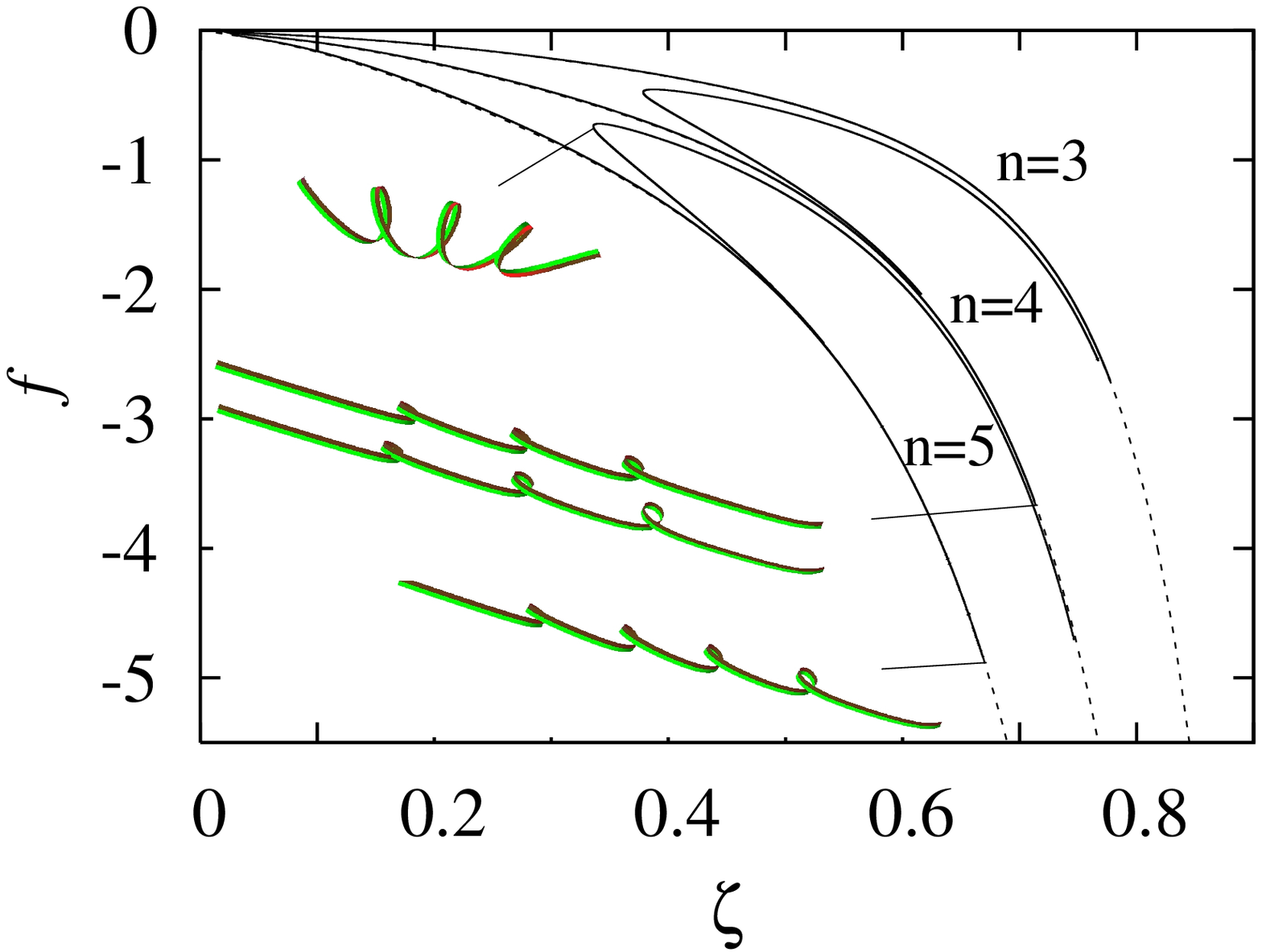}
\caption{\label{fig:small_nu} Force-extension curves for 3-, 4- and 5-turn helices of very low pitch ($\beta_0 = 89^\circ$). {\it Left}: $\nu=0.2$. {\it Right:} $\nu=0.05$. Unlike in previous figures the force here has been normalised by the {\it common} factor $R_0^2=1/(2\pi n_*)^2$ with $n_*=5$. Dashed curves are for the planar elastica for $n=2$ (resp. 3) to 5. For small $\zeta$ these curves have $\left. \partial_\zeta f \right|_{\zeta=\zeta_0}=-\frac{4}{3}(n/n_*)^2$.}
\end{figure}

Figs~\ref{fig:h5shapes}--\ref{fig:small_nu} show results of numerical calculations in which we solved Eqs~(\ref{eqs_vector}) and (\ref{eqs_scalar}) for a finite-length and zero-width ribbon of $n$ helical turns subject to zero-moment ($\bm{M}=\bm{0}$) boundary conditions (thus emulating the conditions in the experiments in \cite{Smith01}). Fig.~\ref{fig:h5shapes} gives the response curve and superimposed ribbon shapes showing non-uniform uncoiling. Note that $\zeta$ is now the end-to-end distance. For extensions less than about 0.8 the curve closely follows the (dashed) helical curve. At larger extensions more complicated hysteresis behaviour is found, with jumps occurring not only under controlled force (known from \cite{Kessler03}), but also under controlled extension. This multistability allows one again to construct multi-phase solutions. Such solutions will in general not be smooth at the interface and, in order to be supported mechanically, would require special joints. Smooth solutions may however exist for certain parameters and boundary conditions.

Fig.~\ref{fig:h37} compares results for various values of $n$ and $\beta_0$ showing that non-monotonic force response is only observed for sufficiently large $\beta_0$, a result also found for the Frenet rod model \cite{Kessler03}. The integer $n$ controls the number of `bumps'. As $n\to\infty$ the helical curve is approached. Fig.~\ref{fig:h37}{\it Left} also shows, through the slopes of the curves at small $\zeta$, that a helix is stiffer in tension than a real ribbon that is allowed to deform from its helical state, as is to be expected given the helical constraint. For comparison, Fig.~\ref{fig:h108} gives results for the Frenet rod. At large $\zeta$ the force response is quite different from that of the inextensible strip, which shows larger oscillations.

Fig.~\ref{fig:small_nu} shows curves at a pitch angle close to $0^\circ$ (at which value the centreline of the ribbon becomes planar and is described by Euler's elastica with intrinsic curvature). An interesting unlooping scenario is observed near this limit with the first response of the helix being a shearing to a (near) planar shape. The subsequent force-response is then essentially that of the planar multi-looped noninflectional elastica with zero end moment whose response curves are included in dashed lines (the remarkable agreement worsens slightly for larger $n$ since the loops are not exactly in one plane). 3D unlooping transitions (again with hysteresis cycles) occur in which successive loops are lost one by one. For smaller Poisson ratio the hockles get tighter. No self-intersections are encountered in all our computations with $w=0$.


We have explained the tension-induced multistability of helical ribbons by means of a geometrically-exact model of an inextensible strip for which we derived new equilibrium equations. Two different types of non-monotonic force-extension behaviour have been found, one approximated by the unwinding of an exact helix, the other by transitions between curves of the looped planar Euler elastica. Unlike in previous rod models we observe this behaviour without the need to choose very large torsional-to-bending stiffness ratios, which are hard to justify in the frame of continuum elasticity theory. In our model the torsional stiffness arises naturally as a consequence of the geometric constraint of developability of the ribbon's surface on account of the crystalline structure of the material. The predictions of this model could conceivably be verified in the controlled environment of nanoribbons \cite{Gao05}. Alternative boundary conditions could easily be accommodated, as could a finite width $w$, while the model could be extended to allow for material anisotropy. The multiple hysteresis behaviour in helical ribbons could be exploited in the design of force probes or bi- or multistable nanoswitches \cite{Rueckes00}.

\begin{acknowledgments}
This work was supported by the UK's Engineering and Physical Sciences Research Council under grant number GR/T22926/01.
\end{acknowledgments}

\bibliography{star}

\end{document}